\documentclass{eas}
\usepackage{graphicx}
\newcommand{\dg}{$^{\circ}$\,}
\newcommand{\as}{$''$\,}
\newcommand{\am}{$'$\,}
\newcommand{\Kdark}{$K_{dark}$\,}
\def\mum{$\mu {\rm m}$}

\begin{document}

%%-----------------------------
%%      the top matter
%%-----------------------------
\title{A SiC TMA GLAO design for PLT?}
\runningtitle{A SiC TMA GLAO design for PLT?}

\author{W. Saunders}\address{School of Physics, University of New South Wales, Sydney, NSW 2052 \email{will@aao.gov.au}}
\author{J.S. Lawrence}\sameaddress{1}
\author{ J.W.V. Storey}\sameaddress{1}
\author{R.Haynes}\address{Anglo-Australian Observatory, Epping, NSW 1710, Australia}
\begin{abstract}
 A `PILOT-Like Telescope' is likely to have differences in science goals to the original PILOT. Furthermore, our understanding of the environmental conditions at Dome~C has changed significantly since the start of the PILOT design study in June 2007. Therefore, it is timely to re-examine some of the basic design decisions. We present here one alternative concept:  a silicon-carbide, GLAO-assisted, three-mirror anastigmat, and possibly equatorial, PILOT-Like-Telescope.
\end{abstract}
\maketitle
%%-----------------------------
%%      your text
%%-----------------------------

\section{Introduction}
The optical design for PILOT arising from the NCRIS design study is for a general purpose, 2.5m optical/IR Ritchey-Chr\'{e}tien/Nasmyth telescope. The optics are diffraction-limited over degree-sized fields, and the delivered image quality is intended to be limited only by free seeing and diffraction. The telescope is to be mounted in an actively ventilated dome to minimise dome and mirror seeing, and to prevent frosting. The telescope and dome are to be mounted on a 30m tower so as to be above the boundary layer turbulence most of the time. The design includes a fast tip-tilt secondary to correct for windshake and also residual boundary layer turbulence above the telescope.

\section{Limitations of PILOT}
The PILOT design has a few undesirable features. These include: the need for a deployable baffle for optical use; the impractibility of simultaneous optical/NIR use; the very large size of the NIR camera (because of the cold stop); the extreme asphericity of the optical corrector lens; the difficulty of mounting all the instruments at the same time. However, the biggest technical issues are environmental:

1. The PILOT design assumed 0.2\dg/hr temperature variation, based on the best available data at the time. However, very rapid temperature variations---two orders of magnitude greater---were discovered by the Givre experiment (Durand \etal\ 2008) and the sonic anemometer data of Travouillon \etal\ (2008). The consequence is significant dome seeing, often 0.5\as (Gillingham 2008).

2. The superb free seeing measured by Lawrence \etal\ (2004), which was based on only 6 weeks data, has not been fully confirmed. The case for wide field optical imaging is very strongly dependent on the seeing (Lawrence \etal\ 2009).

3. The optical sky brightness is still uncertain. An accompanying paper (Saunders \etal\ 2009) suggests that auroral emission and airglow are both significant contributors to the sky brightness at Dome~C, further reducing the attractiveness of wide-field optical work.

There are still significant unresolved site testing issues. There are still no data on the NIR sky brightness at Dome C and, at \Kdark, even the dominant emission mechanism is unknown; the winter-time MIR sky brightness and uniformity is also unmeasured; there is no reason for the boundary layer turbulence to have a Kolmogorov spectrum (with very large implications for adaptive optics); active ventilation of the dome requires knowledge of the humidity at all heights.

\section{An infra-red only telescope}
There is an emerging consensus that any `Pilot-Like Telescope' arising from the ARENA process is likely to be an infrared-only telescope. This changes some of the design considerations; the wavefront error becomes less crucial, so the quality of the optics, dome seeing etc, can be reduced. However, the number of warm mirrors becomes very important, as does the operating temperature of the telescope itself.

For the primary and secondary mirrors, PILOT used light-weighted Zerodur to achieve the required superb image quality. However, for an infrared-only telescope, silicon carbide mirrors (almost certainly made from fused segments) would be an attractive design. The weight would be a factor of two lower, and the thermal responsivity an order of magnitude better. These huge gains would likely recoup the much greater cost of the mirror itself.

\section{A 3-mirror anastigmat design}
JDEM and EUCLID use 3-mirror anastigmat (TMA) designs, which allow diffraction-limited imaging over degree-sized fields with flat image planes, with 3 conics and no further optics. Basic geometry demands an additional fold mirror, and obscuration of the pupil or image plane. TMAs can have an accessible exit pupil; this is invaluable for infrared use, since we can put a cold stop there and then have much simpler cameras. These designs have a very large, but annular, field of view. Any reasonable number of NIR detectors can be accommodated, but the required isoplanatic angles to take proper advantage of a given detector area are larger than for a filled field.

The simplest design is a folded-Cassegrain type, shown in Figure 1(a).  It has a CaF$_2$ lens acting as a window for the large dewar, and it allows simultaneous optical/NIR use via an annular dichroic. With this design, we could have an alt-azimuth telescope (and rotate the entire Cassegrain assembly to correct for image rotation), or go to an equatorial telescope. There is room for a ring of up to 16 HAWAII 4RG-15 detectors around a 1.25\dg field. The optics work best with a very fast primary, $\sim$f/1.25, and an overall speed about f/8. The imaging is diffraction-limited beyond 0.5\mum. The weight of the Cassegrain assembly would be about 500kg for dewar and optics, less than the weight saved by going from Zerodur to silicon carbide.

Going to a Nasmyth design (Figure 1(b)) requires a a large (300mm) CaF2 dewar window, but it can be as thick as needed for adequate strength. The design allows for the use of 8 HAWAII 4RG-15 detectors around a 0.75\dg field. Image rotation is an issue, requiring either rotating the entire instrument; or having image rotators inside the dewar, or moving to an equatorial design. The Nasmyth layout leaves the other Nasmyth focal station free for a MIR instrument.

\section{Towers, tip-tilt and GLAO}
PILOT was designed not to depend on adaptive optics, as AO with multiple guide stars introduces an extra level of complexity, risk and cost. The PILOT design did include rapid tip-tilt correction via M2, using a constellation of guide stars around the field, and (in the NIR) small sub-windows of the main detector array. Simulations showed that this should work well in all NIR broad bands, and at all Galactic latitudes (Saunders 2008). It is a very simple, cheap and robust design, giving remarkably uniform image quality across the 10\am field. It should give diffraction-limited image quality over the whole field at \Kdark when the seeing is better than $\sim 0.5$\as, and at M band in 1\as seeing.

The gains achievable with a GLAO system, with and without a tower, are quantified by Travouillon \etal\ (2009). The simplest GLAO implementation would be to pick off the optical light with a dichroic mirror, and have above and below-focus images on a fast read-out optical detectors (L3V or heavily windowed standard CCDs). Low-order AO should be particularly effective at correcting for the temperature and wind shear arising at the dome aperture.

The optical arm of the TMA design was originally intended for a large optical imaging camera, but could easily be used for an adaptive optics module. The advantage of this strategy would be that the adaptive optics module could be added or removed entirely independently of the NIR camera, and the telescope would remain viable without it, with the NIR camera doing its own tip-tilt guiding in reasonable (seeing $<$1\as) conditions. In poor seeing, MIR work could be undertaken. 
In the TMA design, M2 is conjugate to 15m below M1, close enough to the optimal height to allow it to be used for GLAO correction for wide fields. The large mirror inside the dewar is conjugate to a height $\sim$200m, potentially allowing MCAO use.

The optimal tower height is still to be determined. From the Givre results, it seems that the temperature is much more stable at 10--12m than at 30m, and of course it is always colder. From the results of Aristidi \etal\ (2009), it also seems that at 18--20m, the fraction of time with good seeing is not much worse than at 30m. It seems (Durand {\it p.com}) that the air in the lowest $\sim$ 0.4m above the ice is always very cold and dry, so providing a source of dry air for the dome-ventilation, whatever the tower height. Aristidi \etal\ have shown that the principle effect of lowering the telescope is to reduce the fraction of the time when good seeing is available, but that equally good (or bad) seeing can be achieved at any height. Therefore any decision about tower heights and GLAO necessarily involves a careful tradeoffs between the different uses of the telescope with their different imaging requirements.

\begin{figure}
\begin{center}
\includegraphics[width=9.5cm]{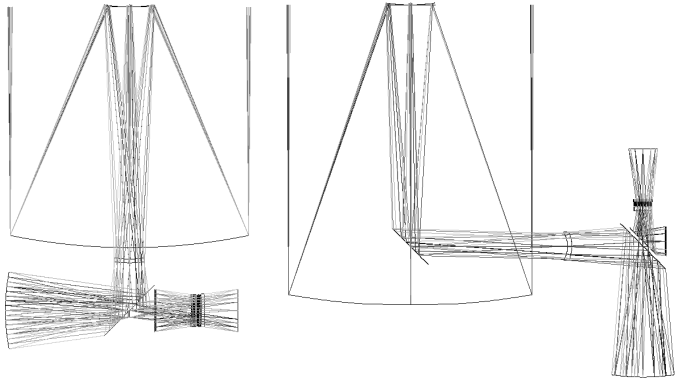}
\end{center}
\caption{TMA layouts for PLT. The entire Cassegrain or Nasmyth assembly is in a dewar in each case. The cold stop is where the beam passes back through the last fold mirror. The camera allows simultaneous NIR/optical use, allowing GLAO with minimal further optics.}
\end{figure}

%%-----------------------------

\end{document}